\documentstyle[12pt]{article}
\renewcommand{\theequation}{\arabic{equation}}
\newcommand{\EQ}{\begin{equation}}
\newcommand{\sh}{\sinh}

\newcommand{\EN}{\end{equation}}

\newcommand{\ket}[1]{\left|#1\right\rangle}      

\newcommand{\bear}{\begin{eqnarray}}
\newcommand{\ear}{\end{eqnarray}}

\begin{document}

\topmargin 0pt
\oddsidemargin 5mm
\newcommand{\NP}[1]{Nucl.\ Phys.\ {\bf #1}}
\newcommand{\PL}[1]{Phys.\ Lett.\ {\bf #1}}
\newcommand{\NC}[1]{Nuovo Cimento {\bf #1}}
\newcommand{\CMP}[1]{Comm.\ Math.\ Phys.\ {\bf #1}}
\newcommand{\PR}[1]{Phys.\ Rev.\ {\bf #1}}
\newcommand{\PRL}[1]{Phys.\ Rev.\ Lett.\ {\bf #1}}
\newcommand{\MPL}[1]{Mod.\ Phys.\ Lett.\ {\bf #1}}
\newcommand{\JETP}[1]{Sov.\ Phys.\ JETP {\bf #1}}
\newcommand{\TMP}[1]{Teor.\ Mat.\ Fiz.\ {\bf #1}}
     
\renewcommand{\thefootnote}{\fnsymbol{footnote}}
     
\newpage
\setcounter{page}{0}
\begin{titlepage}     
\begin{flushright}
IASSNS-HEP-00/15
\end{flushright}
\vspace{0.5cm}
\begin{center}
\large{  Integrability of the $D^2_n$ vertex models with open boundary  } \\
\vspace{1cm}
\vspace{1cm}
 {\large M.J. Martins$^{1,2}$  and X.-W. Guan$^{3}$} \\
\vspace{1cm}
\centerline{\em ${}^{1}$ School of Natural Sciences, Institute for Advanced Study}
\centerline{\em Olden Lane, Princeton, NJ 08540, USA}
\centerline{\em ${}^{2}$ Departamento de F\'isica, Universidade Federal de S\~ao Carlos}
\centerline{\em Caixa Postal 676, 13565-905, S\~ao Carlos, Brazil}
\centerline{\em ${}^{3}$ Institut f\"ur Physik, Technische
  Universit\"at,}
\centerline{\em D-09107 Chemnitz, Germany}
\vspace{1.2cm}   
\end{center} 
\begin{abstract}
We investigate various aspects of the integrability of the vertex models associated
to the $D_n^2$ affine Lie algebra with open boundaries. We first study the solutions
of the corresponding reflection equation compatible with the minimal symmetry
of this system. 
We find three classes of general
solutions, one diagonal solution 
and two non-diagonal families with a free parameter. Next we perform the Bethe ansatz analysis
for some of the associated open $D_2^2$ spin chains and we identify the
boundary having quantum
group invariance. We also discuss a new $D_2^2$ $R$-matrix.

\end{abstract}
\vspace{.2cm}
\vspace{.2cm}
\centerline{February 2000}
\end{titlepage}

\renewcommand{\thefootnote}{\arabic{footnote}}
\section{Introduction}

Much work has been done in integrable lattice statistical mechanics models with
open boundary conditions, since Sklyanin \cite{SK} generalized the quantum inverse 
scattering method to tackle the boundary problem. The bulk Boltzmann weights
of an exactly solvable lattice system are usually the non-null matrix 
elements of a $R$-matrix $R(\lambda)$ which satisfies the Yang-Baxter equation. 
The integrability at boundary, for a given bulk theory, is governed by the
reflection equation, which reads
\begin{equation}
R_{12}(\lambda-\mu ) \stackrel{1}{K}_{-}(\lambda ) R_{21}(\lambda+\mu)
 \stackrel{2}{K}_{-}(\mu ) =  
 \stackrel{2}{K}_{-}(\mu )
 R_{12}(\lambda +\mu ) \stackrel{1}{K}_{-}(\lambda ) R_{21}(\lambda -\mu)
\label{RE1}
\end{equation}
where the matrix $K_-(\lambda)$ describes the reflection at one of the ends of an open chain.
Similar equation should also hold for the reflection $K_+(\lambda )$ at the opposite boundary.
However, for several relevant lattice models $K_+(\lambda )$ can be directly obtained
from $K_-(\lambda )$. For example, this is the case of models
whose $R(\lambda)$ matrix satisfies extra properties such as unitarity,
$P$ and $T$ invariances and crossing symmetry \cite{SK,ME}. 

Therefore, the first step toward
constructing integrable models with open boundaries is to search for solutions of the
reflection equation.
To date, solutions of this equation have been found for a number of lattice
models ranging from vertex systems based on Lie algebras \cite{MEI,DE,BA,IN}
to solid-on-solid models and their restriction \cite{SOS}. Classification of
such solutions for particular systems \cite{CLA} as well as extensions to include
supersymmetric models \cite{SU} can also be found in the literature.

In spite of all these works, there is an interesting vertex model based on the
non-exceptional $D^2_n$ Lie algebra for which little is known about the
solution of the corresponding reflection equation. This is probably related to the fact that
the $D^2_n$ $R$-matrix does not commute for different values of the rapidity \cite{JI},
consequently the trivial diagonal solution $K_-(\lambda)=I$ does not hold for this system
\cite{MEI}. The purpose of this paper is to bridge this gap, by presenting what
we hope to be the minimal solution of the reflection equation for $D^2_n$ vertex models.
This result offers us the possibility to understand a relevant open problem which is 
the integrability  
of the $D^2_n$ vertex model  with quantum algebra symmetry.
In fact,  this symmetry has been found for all vertex models based on non-exceptional
Lie algebras \cite{MEI} except for the $D_n^2$ model. It turns out that,
by carring out a Bethe ansatz analysis, we are able to identify this symmetry
for the simplest $D_2^2$ model and conjecture it for arbitrary values of $n$.

We have organized this paper as follows. We start next section by considering the 
reflection equation for the $D^2_2$ vertex model. We find one diagonal
solution without free parameters and two non-diagonal families which depend on a free
parameter. We also derive the corresponding integrable one-dimensional open
spin chains. 
In section 3 we present the Bethe ansatz solutions of the open $D_2^2$ spin chain
associated to the diagonal $K$-matrix and to a special
manifold of the first non-diagonal family. This allows us to identify the quantum group symmetry for the
$D_2^2$ model.
In section 4 we generalize the $K$-matrices results of section 2 for arbitrary values of $n > 2$.
Section 5 is reserved for our conclusions as well as a discussion on possible new
$D^2_n$ $R$-matrices.
In Appendix A we collect some useful relations and Appendix B contains
a new $D^2_2 $ $R$-matrix as well as its boundary behaviour.

\section{The $D^2_2$ $K$-matrices}

The $D_2^2$ vertex model has four independent degrees of freedom per bond and 
its Boltzmann weights preserve only one $U(1)$ symmetry out of two possible ones.
Here we are interested in looking at solutions of the reflection equation that commute
with this symmetry. We find that the most general $K$-matrix having this property
is 
\begin{equation}
K_-(\lambda)=\left(\matrix{Y_1(\lambda )&0&0&0\cr
0&Y_2(\lambda )&Y_5(\lambda )&0\cr 0&Y_6(\lambda )&Y_3(\lambda )&0 \cr
0&0&0&Y_4(\lambda )\cr }\right)
\end{equation}

Our next step is to substitute this ansatz in equation (\ref{RE1}) and look for
relations that constraint the unknown elements $Y_j(\lambda ), j=1,\ldots , 6$.
Although we have many functional equations, a few of them are actually independent,
and the most suitable ones have been collected in Appendix A. The basic idea
is to try to solve such equations algebraically, which hopefully will produce a
general ansatz for functions $Y_j(\lambda )$ containing several arbitrary
parameters. The general strategy we use is to separate these equations in terms
of ratio of functions depending either on 
$\lambda$ or on $\mu$. From the relations (A.5-A.7) one easly concludes that the simplest
possible solution is to take 
$Y_5(\lambda )=Y_6(\lambda )=0$. This is the diagonal solution, and by employing the
``separation variable method'' described above for the relations (A.5-A.7) we  are able
to fix the following ratios  
\begin{equation}
\frac{Y_2(\lambda )}{Y_1(\lambda )} =
\frac{e^{\lambda }-\beta_1}{e^{-\lambda }-\beta_1},~~ 
\frac{Y_3(\lambda )}{Y_1(\lambda )}= 
\frac{e^{\lambda }-\beta_2}{e^{-\lambda }-\beta_2},~~
\frac{Y_4(\lambda )}{Y_3(\lambda )} =
\frac{e^{\lambda }-\beta_3}{e^{-\lambda }-\beta_3}
\end{equation}
where $\beta_j,~j=1,2,3$ are arbitrary constants. 

These relations enable us to
write an ansatz for three unknown functions in terms of a normalizing factor,
say $Y_1(\lambda )$. Substituting the relations (3)
back to the reflection equation (\ref{RE1}), we conclude that all the parameters
$\beta_j$ are fixed by
\begin{equation}
\beta_1=-\beta_2=-1/\beta_3=\frac{I}{\sqrt{q} }
\end{equation}
where $q$ is the deformation parameter of the $D^2_2$ $R$-matrix \cite{JI}.
This leads us to our first solutions with no free parameter,
\begin{equation}
Y_1^{(1)}(\lambda ) = 1,~~
Y_2^{(1)}(\lambda )  = \frac{e^{\lambda }-\frac{I}{\sqrt{q}}}{e^{-\lambda }-\frac{I}{\sqrt{q}}},~~
Y_3^{(1)}(\lambda )  = \frac{e^{\lambda }+\frac{I}{\sqrt{q}}}{e^{-\lambda }+\frac{I}{\sqrt{q}}},~~
Y_4^{(1)}(\lambda ) =  \frac{e^{\lambda }+\frac{I}{\sqrt{q}}}{e^{-\lambda }+\frac{I}{\sqrt{q}}}
\frac{e^{\lambda }-I\sqrt{q}} {e^{-\lambda }-I\sqrt{q} }
\end{equation}

Next we turn our search for non-diagonal solutions now with both $Y_5(\lambda )$
and $Y_6(\lambda )$ non null. From the equations (A.8-A.11), we notice that it is
possible to solve $Y_2(\lambda ),Y_3(\lambda )$ and $Y_6(\lambda )$ in terms
of $Y_5(\lambda )$. At this point we should keep in mind that we are looking
for regular $K$-matrices, i.e. $K_-(0)\sim $ identity.
After some simplifications, we find the following general solutions
\begin{equation}
\epsilon_1(1+e^{2\lambda })[Y_5(\lambda )+Y_6(\lambda )]= 
\epsilon_2 e^{\lambda }[Y_5(\lambda )-Y_6(\lambda )] 
\end{equation}
\begin{equation}
(e^{2\lambda }-1)[Y_2(\lambda )+Y_3(\lambda )]  = 
\epsilon_3 e^{\lambda }[Y_6(\lambda )-Y_5(\lambda )]
\end{equation}
\begin{equation}
Y_2(\lambda )-Y_3(\lambda )  =  \epsilon_4 [Y_5(\lambda )-Y_6(\lambda )]
\end{equation}
where $\epsilon_j$  are four arbitrary parameters. These are linear
equations which can be easily solved for the ratios $Y_2(\lambda )/Y_5(\lambda ),
Y_3(\lambda )/Y_5(\lambda )$ and $Y_6(\lambda )/Y_5(\lambda )$. Taking this into
account as well as equations (A.5) and (A.7), we end up with the following
ansatz for functions $Y_j(\lambda )$
\begin{equation}
Y_1^{}(\lambda )  =  (\epsilon_5e^{2\lambda}+\epsilon_6e^{\lambda}+\epsilon_7)/e^{\lambda},~~
Y_2^{}(\lambda )  =  (1+e^{2\lambda})\left[\epsilon_4(e^{2\lambda }-1)-\epsilon_3e^{\lambda}\right]
\end{equation}
\begin{equation}
Y_3^{}(\lambda )  =  (1+e^{2\lambda})\left[-\epsilon_4(e^{2\lambda }-1)-\epsilon_3e^{\lambda}\right],~~
Y_4^{}(\lambda )  =  (\epsilon_8e^{2\lambda}+\epsilon_9e^{\lambda}+\epsilon_{10})e^{\lambda}
\end{equation}
\begin{equation}
Y_5^{}(\lambda )  =  (e^{2\lambda}-1)\left[\epsilon_2e^{\lambda }+\epsilon_1(1+e^{2\lambda})\right],~~
Y_6^{}(\lambda )  =  (e^{2\lambda}-1)\left[\epsilon_2e^{\lambda }-\epsilon_1(1+e^{2\lambda})\right]
\end{equation}
having altogether ten free parameters. Substituting this ansatz
back to the reflection equation and after involving algebraic manipulations, we find that
nine parameters are in fact fixed, leading us to two classes of non-diagonal
solution with a free parameter. The first class is given by
\begin{equation}
Y_1^{(2)}(\lambda,\xi_-)  =  (e^{2\lambda }+q)(\xi_-^2 q e^{2\lambda }-1 )e^{-\lambda},~~
Y_4^{(2)}(\lambda,\xi_-)  =  (e^{2\lambda }+q)(\xi_-^2 q- e^{2\lambda })e^{\lambda }
\end{equation}
\begin{equation}
Y_2^{(2)}(\lambda,\xi_-)  =  \frac{(1+e^{2\lambda })}{2}\left[2(e^{2\lambda }-1)\xi _-q-e^{\lambda }(1+q)(1-\xi _-^2q)\right]
\end{equation}
\begin{equation}
Y_3^{(2)}(\lambda,\xi_-)  =  \frac{(1+e^{2\lambda })}{2}\left[-2(e^{2\lambda }-1)\xi _-q-e^{\lambda }(1+q)(1-\xi _-^2q)\right]
\end{equation}
\begin{equation}
Y_5^{(2)}(\lambda,\xi_-)=Y_6^{(2)}(\lambda,\xi_{-})  =  \frac{(e^{2\lambda }-1)}{2}(1-q)(\xi _-^2q+1)e^{\lambda }
\end{equation}
while the second family is
\begin{equation}
Y_1^{(3)}(\lambda,\xi_-)  =  (e^{2\lambda }-q)(\xi_{-}e^{2\lambda }-1 )e^{-\lambda},~~
Y_4^{(3)}(\lambda,\xi_-)  =  (e^{2\lambda }-q)(\xi_{-}-e^{2\lambda })e^{\lambda }
\end{equation}
\begin{equation}
Y_2^{(3)}(\lambda,\xi_-)=Y_3^{(3)}(\lambda,\xi_{-})  =  \frac{(1+e^{2\lambda })}{2}(1-q)(\xi_{-}-1)e^{\lambda }
\end{equation}
\begin{equation}
Y_5^{(3)}(\lambda,\xi_-)  =   \frac{(e^{2\lambda }-1)}{2}\left[2(e^{2\lambda }+1)\sqrt{\xi _{-}q}+(1+q)(1+\xi _-)e^{\lambda }\right]
\end{equation}
\begin{equation}
Y_6^{(3)}(\lambda,\xi_-)  =   \frac{(e^{2\lambda }-1)}{2}\left[-2(e^{2\lambda }+1)\sqrt{\xi _-q}+(1+q)(1+\xi _-)e^{\lambda }\right]
\end{equation}
where $\xi_{-} $ is an arbitrary parameter. 

Since the $D^2_2$ $R$-matrix is $PT$ invariant and crossing symmetric, the 
$K_+(\lambda )$ matrices at the 
opposite boundary are easily derived from the above solutions
\cite{ME,MEI}. More precisely, we have
\begin{equation}
K_+(\lambda, \xi _+ )=K_-^t(\ln[q]-\lambda,\xi _+)M
\end{equation}
where $M$ is a matrix related to
the crossing matrix $V$ by $M= V^{t} V$ \cite{MEI}. From the results
of  Appendix A, we have that for the $D^2_2$
model $M$ is given by
\begin{equation}
M={\rm diag}(q,1,1,q^{-1})
\end{equation}

Having found the $K_{\pm }(\lambda )$ matrices, one can construct the 
corresponding commuting transfer matrix $\tau(\lambda )$. Following Sklyanin 
\cite{SK}, we have
\begin{equation}
t^{(l,m)}(\lambda )={\rm Tr}_a\left[\stackrel{a}{K}_+^{(m)}(\lambda )T(\lambda )\stackrel{a}{K}^{(l)}_-(\lambda )T^{-1}(-\lambda )\right]
\end{equation}
where
$T(\lambda )=R_{aL}(\lambda )\cdots R_{a1}(\lambda )$ is the monodromy matrix 
of the associated closed chain with $L$ sites.  This means that the three families
of $K_-(\lambda )$ matrices we found will produce nine possible types of open boundary
conditions. The corresponding Hamiltonian of the spin chains with open boundaries
are obtained by
expanding the transfer matrix $t^{(l,m)}(\lambda )$ in powers of $\lambda $. 
When ${\rm Tr}[K_+^{(m)}(0)]$ is non-null, the Hamiltonian $H^{(l,m)}$ is proportional
to the first-order expansion \cite{SK}
\begin{equation}
H^{(l,m)}=\sum _{k=1}^{L-1}H_{k,k+1}+\frac{1}{2\zeta }\frac{d\stackrel{a}{K}_-^{(l)}(\lambda )}{d\lambda }|_{\lambda =0}+
\frac{{\rm Tr}_a\left[\stackrel{a}{K}_-^{(m)}(0)H_{La}\right]}{{\rm Tr}\left[K_+^{(m)}(0)\right]}
\end{equation}
where $H_{k,k+1}=P_{k,k+1}\frac{d}{ d \lambda }R_{k,k+1}(\lambda )|_{\lambda =0}$ is
the two-body bulk Hamiltonian and $\zeta $ is the normalization $R_{12}(0)=\zeta P_{12}$ 
\footnote{The normalization we use for $R(\lambda )$ (see Appendix A)  produces
$\zeta =(q-1/q)^2$ for $D^2_2$ model.}

For  the first two solutions we indeed have ${\rm Tr}[K_+(0)]\neq 0$ while for
the third one
${\rm Tr}[K_+(0)]= 0$. In this last case
one has to consider the second order expansion in the spectral parameter $\lambda $
\cite{HH}. We find convenient to write the expression for the Hamiltonians in terms  of
Pauli matrices $\sigma_{\alpha,i}^{\pm}$ and $\sigma_{\alpha,i}^{z} $ with components $\alpha=\uparrow,\downarrow$
acting on the site $i$ of a lattice of size $L$. In terms of these operators 
and up to irrelevant additive constants\footnote{We also note that we have normalized the Hamiltonian by the
pure imaginary number.}, we have
\begin{eqnarray}
H^{(l,m)} & = & -\frac{I(q-1/q)}{2}
\sum _{k=1}^{L-1}{\tilde{H}}_{k,k+1}+I\frac{(q-1/q)^2}{2}\left\{ \right.\nonumber\\
& &
\left. \sum_{\alpha=\uparrow,\downarrow}\mu_{\alpha}^{(l)}(\xi_-)\sigma_{\alpha,1}^z
+\delta^{(l)}\sigma_{\uparrow,1}^z \sigma_{\downarrow,1}^z+J_{\uparrow}^{(l)}(\xi_-)\sigma_{\uparrow,1}^+\sigma_{\downarrow,1}^-
+J_{\downarrow}^{(l)}(\xi_{-})\sigma_{\downarrow,1}^+\sigma_{\uparrow,1}^-\right.\nonumber\\
& &
\left. -\sum_{\alpha=\uparrow,\downarrow}\mu_{\alpha}^{(m)}(\xi_+)\sigma_{\alpha,L}^z
+\delta^{(m)}\sigma_{\uparrow,L}^z \sigma_{\downarrow,L}^z+J_{\downarrow}^{(m)}(\xi_+)\sigma_{\uparrow,L}^+\sigma_{\downarrow,L}^-
+J_{\uparrow}^{(m)}(\xi_{+})\sigma_{\downarrow,L}^+\sigma_{\uparrow,L}^-\right\}
\nonumber\\
\end{eqnarray}
where the expression of the 
bulk part ${\tilde{H}}_{k,k+1}$ is 
\begin{eqnarray}
{\tilde{H}}_{k,k+1} & = &
\frac{(q-1/q)}{2}\left[(\sigma_{\uparrow,k}^z+\sigma_{\downarrow,k}^z)(\sigma_{\uparrow,k+1}^{+}
\sigma_{\downarrow,k+1}^{-}+\sigma_{\uparrow,k+1}^{-}\sigma_{\downarrow,k+1}^{+})
-(\sigma_{\uparrow,k}^{+}\sigma_{\downarrow,k}^{-}+
\sigma_{\uparrow,k}^{-}\sigma_{\downarrow,k}^{+})
(\sigma_{\uparrow,k+1}^z+\sigma_{\downarrow,k+1}^z)\right]\nonumber \\
& &
+(\sqrt{q}-\frac{1}{\sqrt{q}})^2\left[\sigma^{+}_{\uparrow,k}\sigma_{\uparrow,k+1}^{-}
\sigma_{\downarrow,k}^{-}\sigma_{\downarrow,k+1}^{+}+
\sigma^{-}_{\uparrow,k}\sigma_{\uparrow,k+1}^{+}
\sigma_{\downarrow,k}^{+}\sigma_{\downarrow,k+1}^{-} \right. \nonumber\\
& &
\left. +\sigma^{+}_{\uparrow,k}\sigma_{\uparrow,k+1}^{+}
\sigma_{\downarrow,k}^{-}\sigma_{\downarrow,k+1}^{-}+
\sigma^{-}_{\uparrow,k}\sigma_{\uparrow,k+1}^{-}
\sigma_{\downarrow,k}^{+}\sigma_{\downarrow,k+1}^{+}
\right]\nonumber \\
& &
-2\left[(\sigma^+_{\uparrow,k}\sigma^{-}_{\uparrow,k+1}+
\sigma_{\uparrow,k}^{-}\sigma^{+}_{\uparrow,k+1})(1+
\sigma^{z}_{\downarrow,k} \sigma^{z}_{\downarrow,k+1})
+(\sigma^{+}_{\downarrow,k} \sigma^{-}_{\downarrow,k+1}
+\sigma^{-}_{\downarrow,k}\sigma^{+}_{\downarrow,k+1})(1+\sigma^{z}_{\uparrow,k}
\sigma^{z}_{\uparrow,k+1})\right]\nonumber \\
& &
+(\sqrt{q}+\frac{1}{\sqrt{q}})\left[(\sigma^+_{\uparrow,k}\sigma^{-}_{\uparrow,k+1}+
\sigma_{\uparrow,k}^{-}\sigma^{+}_{\uparrow,k+1})(1-
\sigma^{z}_{\downarrow,k} \sigma^{z}_{\downarrow,k+1}) \right. \nonumber\\
& &
 \left. +(\sigma^{+}_{\downarrow,k} \sigma^{-}_{\downarrow,k+1}
+\sigma^{-}_{\downarrow,k}\sigma^{+}_{\downarrow,k+1})(1-\sigma^{z}_{\uparrow,k}
\sigma^{z}_{\uparrow,k+1})\right]\nonumber \\
& &
-(\sqrt{q}-\frac{1}{\sqrt{q}})\left[(\sigma_{\uparrow,k}^{+}\sigma_{\downarrow,k+1}^{-}
+\sigma _{\uparrow,k}^{-}\sigma_{\downarrow,k+1}^{+})(\sigma^z_{\downarrow,k}-\sigma^z_{\uparrow,k+1})
+(\sigma^+_{\downarrow,k}\sigma^{-}_{\uparrow,k+1}+\sigma^-_{\downarrow,k}
\sigma^{+}_{\uparrow,k+1})(\sigma^z_{\uparrow,k}-\sigma^z_{\downarrow,k+1})\right] \nonumber\\
& &
+[1-\frac{(q+1/q)}{2}](\sigma_{\uparrow,k}^{z}\sigma_{\downarrow,k}^{z}+\sigma_{\uparrow,k+1}^{z}
\sigma_{\downarrow,k+1}^{z})
-\frac{(\sqrt{q}-\frac{1}{\sqrt{q}})^2}{4}(\sigma_{\uparrow,k}^{z}\sigma_{\downarrow,k+1}^{z}+
\sigma_{\downarrow,k}^{z}\sigma_{\uparrow,k+1}^{z})\nonumber \\
& &
-[\frac{(q+1/q)}{4}+\frac{3}{2}](\sigma^{z}_{\uparrow,k}\sigma _{\uparrow,k+1}^{z}
+\sigma_{\downarrow,k}^{z}\sigma_{\downarrow,k+1}^{z})
+\frac{(q-1/q)}{2}\sum_{\alpha=\uparrow,\downarrow}(\sigma^{z}_{\alpha,k}-\sigma _{\alpha,k+1}^{z})
- 2(q+\frac{1}{q})I_{k,k+1} \nonumber
\end{eqnarray} 

Turning to the boundary interactions we found that the chemical potentials are given by
\begin{equation}
\mu_{\alpha }^{(l)}(\xi) = 
\left\{\begin{array}{l}
\mu_{\uparrow}^{(1)}(\xi)=1/2-I\frac{\sqrt{q}}{1+q},~~\mu_{\downarrow}^{(1)}(\xi)=1/2+I\frac{\sqrt{q}}{1+q} \\
\mu_{\uparrow}^{(2)}(\xi)=-\frac{(1+q+ 2\xi q)}{(1+q)(\xi ^2q^2-1)},~~ 
\mu_{\downarrow}^{(2)}(\xi)=-\frac{(1+q-2\xi q)}{(1+q)(\xi ^2q^2-1)}\\
\mu_{\uparrow}^{(3)}(\xi)=\mu_{\downarrow}^{(3)}=\frac{1}{1-\xi}
\end{array}\right. 
\end{equation}
while the on-site parameters $\delta^{(l)}$ and $ J_{\alpha}^{(l)}(\xi)$ are 
\begin{equation}
\delta^{(l)} = 
\left\{\begin{array}{l}
\delta^{(1)}=\frac{(q-1)}{2(1+q)} \\
\delta^{(2)}=\frac{(q-1)}{2(1+q)} \\
\delta^{(3)}=\frac{(1+q)}{2(q-1)}
\end{array}\right. 
\end{equation}
and
\begin{equation}
J_{\alpha}^{(l)}(\xi) = 
\left\{\begin{array}{l}
J_{\uparrow}^{(1)}(\xi)=J_{\downarrow}^{(1)}=0 \\
J_{\uparrow}^{(2)}(\xi)=J_{\downarrow}^{(2)}(\xi ) = \frac{(q-1)(1+\xi ^2q)}{(1+q)(\xi ^2q-1)} \\
J_{\uparrow}^{(3)}(\xi )  = \frac{(1+q)(1+\xi)+ 4\sqrt{q\xi } }{(q-1)(\xi -1)},~~
J_{\downarrow}^{(3)}(\xi )  = \frac{(1+q)(1+\xi)- 4\sqrt{q\xi } }{(q-1)(\xi -1)}
\end{array}\right. 
\end{equation}

A natural question to be asked is which (if any) of these solutions would lead us to an integrable $D_2^2$ model
with quantum algebra symmetry. One way to investigate that is by applying the Bethe ansatz  method to diagonalize
the above  open spin chains. This allows us to extract information 
about the eigenspectrum, which in the case of quantum algebra invariance, should be  
highly degenerated (see e.g. \cite{SA}). In next section we will discuss this problem in details.

\section{Bethe ansatz analysis}

The purpose of this section is to study the spectrum of some of the open spin chains presented in section 2
by the coordinate Bethe ansatz formalism. One of our motivations is to identify the boundary that leads us
to the quantum group symmetry. We begin by noticing that the total number of spins
${\hat{N}}_s=\sum_{i=1}^{L} \sum_{\alpha=\uparrow,\downarrow} \sigma_{\alpha,i}^{z}$ is a conserved
quantity and its eigenvalues $ns$ labels the many possible disjoint sectors of the Hilbert space.
Therefore, the wave function solving the eigenvalue problem $ H\ket{\Psi_{ns}}= E^{(l,m)}(L) \ket{\Psi_{ns}} $
can be written as follows
\begin{equation}
\ket{\Psi_{ns}}= \sum_{\alpha_j} \sum_{x_{Q_j}} f^{(\alpha_1,\cdots,\alpha_n)}(x_{Q_1},\cdots, x_{Q_{ns}}) 
\sigma_{\alpha_1,x_{Q_1}}^{+} \cdots 
\sigma_{\alpha_{ns},x_{Q_{ns}}}^{+}  \ket{0}
\end{equation}
where $\ket{0}$ denotes the ferromagnetic state (all spins up) and 
$1\leq x_{Q_1}\leq x_{Q_2}\leq \cdots \leq x_{\mbox{\scriptsize $
Q_{ns}$}}\leq L$ indicate the positions of the  spins. 

We will start our study by first considering the open spin chain $H^{(1,1)}$ 
corresponding to the diagonal $K$-matrix solution. As it is customary we begin our discussion of the
eigenvalue problem in the sector of one down spin, $ns=1$. In 
this sector, we find that for
$1< x < L$
\begin{equation}
\frac{I}{2(1/q-q)}E^{(1,1)}(L) f^{(\alpha)}(x)= 
(L-2)\Delta f^{(\alpha)}(x) +f^{(\alpha)}(x+1)+f^{(\alpha)}(x-1) +\frac{(1-q)^2}{4q}
f^{(\alpha)}(x),~~\alpha=\uparrow,\downarrow
\end{equation}
where we have defined $\Delta=q+1/q$. 
The matching condition at the left and right boundaries gives us the following constraints
\begin{equation}
\left(\begin{array}{c} f^{(\uparrow)}(0) \\ f^{(\downarrow)}(0) \end{array} \right )   =
\left(\begin{array}{cc} \Delta -p_{1\uparrow} & d_{1\uparrow} \\ 
d_{1\downarrow} & \Delta-p_{1\downarrow}  
\end{array} \right ) 
\left(\begin{array}{c} f^{(\uparrow)}(1) \\ f^{(\downarrow)}(1) \end{array} \right )
\end{equation}
and
\begin{equation}
\left(\begin{array}{c} f^{(\uparrow)}(L+1) \\ f^{(\downarrow)}(L+1) \end{array} \right )   =
\left(\begin{array}{cc} \Delta -p_{L\uparrow} & d_{L\uparrow} \\ 
d_{L\downarrow} & \Delta-p_{L\downarrow}  
\end{array} \right ) 
\left(\begin{array}{c} f^{(\uparrow)}(L) \\ f^{(\downarrow)}(L) \end{array} \right )
\end{equation}
where the matrices parameters are given by
\begin{equation}
p_{1\uparrow}= \frac{3I+\sqrt{q}+q^2(I+3\sqrt{q})}{4q(I+\sqrt{q})},~~
p_{1\downarrow}= \frac{-3I+\sqrt{q}+q^2(-I+3\sqrt{q})}{4q(-I+\sqrt{q})}
\end{equation}
\begin{equation}
p_{L\uparrow}= \frac{3-2q+3q^2-2I\sqrt{q}+2Iq\sqrt{q}}{4q},~~
p_{L\downarrow}= \frac{3-2q+3q^2+2I\sqrt{q}-2Iq\sqrt{q}}{4q}
\end{equation}
\begin{equation}
d_{1\uparrow}=d_{1\downarrow}=-d_{L\uparrow}=-d_{L\downarrow}= \frac{q-1/q}{4}
\end{equation}

In order to go ahead it is crucial to notice 
that both boundary constraints (30) and (31) can be diagonalized
by the $same$ unitary transformation $U$. After performing this transformation the new components 
${\tilde{f}}^{\alpha}(x)=U f^{\alpha}(x)$  satisfy
\begin{equation}
\left(\begin{array}{c} {\tilde{f}}^{(\uparrow)}(0) \\ {\tilde{f}}^{(\downarrow)}(0) \end{array} \right )   =
\left(\begin{array}{cc} 1 & 0 \\ 
0 &  \frac{\Delta}{2}
\end{array} \right ) 
\left(\begin{array}{c} {\tilde{f}}^{(\uparrow)}(1) \\ {\tilde{f}}^{(\downarrow)}(1) \end{array} \right )
\end{equation}
and
\begin{equation}
\left(\begin{array}{c} {\tilde{f}}^{(\uparrow)}(L+1) \\ {\tilde{f}}^{(\downarrow)}(L+1) \end{array} \right )   =
\left(\begin{array}{cc} \frac{\Delta}{2} & 0 \\ 
0 &  1 
\end{array} \right ) 
\left(\begin{array}{c} {\tilde{f}}^{(\uparrow)}(L) \\ {\tilde{f}}^{(\downarrow)}(L) \end{array} \right )
\end{equation}

Clearly, equation (29) for $1<x<L$ remains the same but now for the transformed amplitudes
${\tilde{f}}^{(\alpha)}(x)$.
Now we reached a point in which one can try the usual Bethe ansatz (e.g. see ref. \cite{BET,BET1}), namely
\begin{equation}
{\tilde{f}}^{(\alpha)}(x)= A_{\alpha}(k)e^{ikx}-A_{\alpha}(-k)e^{-ikx}
\end{equation}
and by substituting this ansatz in (29) we obtain the following eigenvalue
\begin{equation}
\frac{I}{2(1/q-q)}E^{(1,1)}(L) = (L-2)\Delta +2\cos(k)  
+\frac{(1-q)^2}{4q}
\end{equation}

The fact that this ansatz should be also valid for the ends $x=1$ and $x=L$ provides us
constraints for the amplitudes $A(k)$ and $A(-k)$, which reads
\begin{equation}
A(-k)=-e^{ik}A(k)~~{\rm and}~~ A(-k)= \frac{(1-\frac{\Delta}{2}e^{-ik})}
{(1-\frac{\Delta}{2}e^{ik})}e^{2i(L+1)k}A(k)
\end{equation}
whose 
compatibility  gives 
a restriction on the momentum $k$, namely 
\begin{equation}
e^{2ikL}\frac{(e^{ik}-\frac{\Delta}{2})}
{(\frac{\Delta}{2}e^{ik}-1)}=1
\end{equation}

The next task is to generalize 
these results for arbitrary numbers of down spins. For a general multiparticle state,
we assume the Bethe ansatz wave function 
\begin{equation}
{\tilde{f}}^{(\alpha_1,\cdots,\alpha_n)}(x_{Q_1},\cdots, x_{Q_{ns}}) =
\sum _{P} \mbox{{\rm sgn}}(P)\prod _{j=1}^{ns}
e^{\mbox{\scriptsize $[ik_{p_j}x_{Q_j}]$}}
A(k_{PQ_1},\cdots, k_{PQ_{N_e} })_{\mbox{\scriptsize 
$\alpha_{Q_1}, \cdots, \alpha_{Q_{ns}}$ }} 
\end{equation}
where $P$ is the sum over all the permutations of the momenta, including the 
negations $k_j \rightarrow -k_j$, and
the symbol $sgn$ 
accounts for the sign of the permutations and negations. 
It turns out that for
configurations  such that
$|x_{Q_i}-x_{Q_j}| \geq 2$ the open 
spin chain $H^{(1,1)}$ behaves as a free theory and the corresponding eigenvalues are
\begin{equation}
\frac{I}{2(1/q-q)}E^{(1,1)}(L) = (L-1)\Delta   +
\sum ^{ns}_{j=1}[2\cos(k_j) -\Delta]
+\frac{(1-q)^2}{4q}
\end{equation}

The new ingredient for $ns \geq 2$ is that the nearest neighbor spin configurations
enforce constraints on the amplitude of the wave function. This condition enhances a relation
between the exchange of two states such as $\{(k_i,\alpha_i); (k_j, \alpha_j)\} $ and
$\{(k_j,\alpha_j); (k_i, \alpha_i)\} $ which ultimately is represented by the two-body
scattering
\begin{equation}
 A_{\cdots \mbox{\scriptsize $ \alpha _j,\alpha _i$}\cdots }(\cdots ,k_j,k_i,\cdots )=
 S_{i,j}(k_i,k_j)A_{\cdots \mbox{\scriptsize $ \alpha_i,\alpha_j$}\cdots }(\cdots ,k_i,k_j,\cdots )
\end{equation}
while the reflection at the left and right ends generalizes equation (29), which now reads
\begin{eqnarray}
A_{\mbox{\scriptsize $ \alpha _i,$}\cdots }(-k_j,\cdots ) & = &
-e^{ik_j}
A_{\mbox{\scriptsize $ \alpha _i,$}\cdots }(k_j,\cdots )\\
A_{\cdots ,\mbox{\scriptsize $ \alpha_i$}}(\cdots, -k_j) & = &
\frac{(1-\frac{\Delta}{2}e^{-ik_j})}
{(1-\frac{\Delta}{2}e^{ik_j})}e^{2i(L+1)k_j}
A_{\cdots \mbox{\scriptsize $ \alpha_i,$}}(\cdots,k_j)
\end{eqnarray}

Fortunately, the bulk two-body scattering amplitude $S_{i,j}(k_i,k_j)$ has been recently
identified in ref.\cite{MAR} for the periodic chain. This result is of enormous help here since
it allows us to choose the suitable parametrization for the momenta $k_j$ in terms of
the $S$-matrix rapidities $\lambda_j$, which is
\begin{equation}
e^{ik_j} = \frac{\sh(\lambda_j -i\gamma/2)}{\sh(\lambda_j +i\gamma/2)}
\end{equation}
where we have conveniently defined $q=e^{i\gamma}$. For explicit expression of the non-null
$S$-matrix elements see ref.\cite{MAR}. 

In this general case, the compatibility between the bulk and boundary scattering constraints
(43-45) leads us to the Bethe ansatz  equation for the momenta $k_j$ 
\begin{equation}
e^{2ik_jL}\frac{(e^{ik_j}-\frac{\Delta}{2})}
{(\frac{\Delta}{2}e^{ik_j}-1)}=
\Lambda_j(k_1,\cdots,k_{ns})
\end{equation}
where $\Lambda_j(k_1,\cdots,k_{ns})$ are the eigenvalues of the auxiliary inhomogeneous transfer
matrix  $t_j=S_{jns}(k_j,k_{ns}) \cdots S_{j1}(k_j,k_1) S_{1j}(k1,-k_j) \cdots S_{nsj}(k_{ns},-k_j)$. 
The integrability of this latter inhomogenous problem follows from 
the fact that the $2 \times 2$ identity $K$-matrix
is
a solution of the reflection equation associated to the two-body scattering $S_{ij}$.
As was shown in ref. \cite{MAR} there is no need of a second Bethe ansatz to solve this
auxiliary eigenvalue problem. By adapting the results of ref.\cite{MAR} to our case and by
relating the momenta $k_j$ and the rapidities $\lambda_j$ by equation (46) we find that
the Bethe ansatz equations are given by
\begin{eqnarray}
\left [ \frac{\sh(\lambda_j-i\gamma/2)}{\sh(\lambda_j +i\gamma/2)} 
\right ]^{2L}  
\frac{\cosh(\lambda_j+i\gamma/2)}
{\cosh(\lambda_j-i\gamma/2)} & =&
 \prod_{k=1}^{ns} \frac{ \sh(\lambda_j/2 - \lambda_k/2 -i\gamma/2)}
{ \sh(\lambda_j/2 - \lambda_k/2 +i\gamma/2)}
\frac{ \sh(\lambda_j/2 + \lambda_k/2 -i\gamma/2)}
{ \sh(\lambda_j/2 + \lambda_k/2 +i\gamma/2)} \nonumber\\
& &
j=1, \cdots, ns 
\end{eqnarray}
and the eigenvalues (42) in terms of the rapidities $\lambda_j$ are
\begin{equation}
E^{(1,1)}(L)= -8 \sin^3(\gamma) \sum_{j=1}^{ns} \frac{1}{\cos(\gamma)-\cosh(2\lambda_j)}
-4(L-1)\sin(2\gamma) + 4\sin(\gamma)\sin^2(\gamma/2)
\end{equation}

Clearly, the Bethe ansatz equations  
for the open spin chain $H^{(1,1)}$  are not just the ``doubling'' of the corresponding
results of the closed chain with periodic boundary conditions \cite{RE,MAR} 
due to an additional boundary left hand factor. We recall here that 
the ``doubling'' property has been
argued \cite{ME3} to be  one of the main features of a  quantum algebra invariant open spin
chain at least for standard forms of comultiplication. Looking at the 
spectrum of $H^{(1,1)}$, however, we notice a certain pattern of degeneracies which suggests
an underlying hidden symmetry. It could be that the diagonal boundary solution corresponds
to an asymmetric form of coproduct since this, in principle, is allowed too \cite{CAR}.

Next we turn our attention to the first non-diagonal solution and its corresponding open spin
chain. In this case, at least for generic values of $\xi_{\pm}$, the Bethe ansatz 
construction we just explained above needs further generalizations. This can be seen even
at the level of one down spin state, since there is not a unique
 transformation that 
diagonalizes both left and right boundary matrix problems. However, there is a particular
manifold, $\xi_{+}=q\xi_{-}$, in which our previous Bethe ansatz formulation is still valid.
Fortunately, as we shall see below, this special manifold will be sufficient to single out
the boundary leading us to the quantum algebra symmetry. Since for $\xi_{+}=q \xi_{-}$,
the Bethe ansatz analysis is very similar to the one just
described above, we restrict ourselves  to present only the final results.
We found that the Bethe ansatz equations for the Hamiltonian $H^{(2,2)}$ at $\xi_{+}=q\xi_{-}$
are 
\begin{equation}
\left [ \frac{\sh(\lambda_j-i\gamma/2)}{\sh(\lambda_j +i\gamma/2)} 
\right ]^{2L} = 
 \prod_{k=1}^{ns} \frac{ \sh(\lambda_j/2 - \lambda_k/2 -i\gamma/2)}
{ \sh(\lambda_j/2 - \lambda_k/2 +i\gamma/2)}
\frac{ \sh(\lambda_j/2 + \lambda_k/2 -i\gamma/2)}
{ \sh(\lambda_j/2 + \lambda_k/2 +i\gamma/2)},~
j=1, \cdots, ns \\
\end{equation}
while the corresponding eigenvalues are given by
\begin{equation}
E^{(1,1)}(L)= -8 \sin^3(\gamma) \sum_{j=1}^{ns} \frac{1}{\cos(\gamma)-\cosh(2\lambda_j)}
-4(L-1)\sin(2\gamma) -4\sin(\gamma)\left [ \sum_{\alpha=\uparrow,\downarrow}(\mu_{\alpha}^{(2)}(\xi_{-})
-\mu_{\alpha}^{(2)}(\xi_+) ) +2 \delta^{(2)} \right ]
\end{equation}

Now the Bethe ansatz equations do have the ``doubling'' property at
$\xi_{+}=q\xi_{-}$ and this is an extra motivation to investigate the eigenspectrum of
$H^{(2,2)}$. It turns out that  at the value $\xi_{-}=0$ and therefore 
$\xi_{+}=0$ we discover that the spectrum of the open chain $H^{(2,2)}$ is  
specially highly degenerated. In fact, after some algebraic manipulations,
we check that for
$\xi_{\pm}=0$ the Hamiltonian 
$H^{(2,2)}$ has the appropriate boundary coefficients to ensure commutation
with $U_q(D_2^2)$. Therefore, we finally managed to identify the quantum algebra symmetry
for the $D_2^2$ vertex model.

Finally, it seems desirable to solve the open spins chains associated to the non-diagonal
solutions for arbitrary values of the parameters $\xi_{\pm}$. The 
coordinate Bethe ansatz method, however, leads us
to cumbersome calculations even for the first excitation over the reference state. In such
general case it seems wise to tackle this problem by
using a more unifying technique such as the
algebraic Bethe ansatz approach. Since the basics of this method has been recently
developed
for the $D_n^2$ vertex models \cite{MAR}  we hope to return to this problem elsewhere.

\section{The $D^2_n$ $K$-matrices}

Here we shall consider the generalizations of the $K$-matrices solutions of section 2 for the general
$D_n^2$ model. This system has $n-1$ distinct $U(1)$ conserved charges, and the $K$-matrix ansatz compatible
with these symmetries can be represented by the following block diagonal matrix
\begin{equation}
K_{-}(\lambda)= {\rm diag}(Y_1(\lambda), \cdots,Y_{n-1}(\lambda), \hat{A}(\lambda), Y_{n+2}(\lambda), \cdots,Y_{2n}(\lambda))
\end{equation}
where $\hat{A}(\lambda)$ is a $ 2 \times 2$ matrix
\begin{equation}
\hat{A}(\lambda)=\left(\matrix{Y_n(\lambda )&Y_{2n+1}(\lambda)\cr
Y_{2n+2}(\lambda )&Y_{n+1}(\lambda )\cr
}\right)
\end{equation}
where $Y_j(\lambda),~j=1,\cdots,2n+2$ are functions we have determined by solving the reflection 
equation. Notice that for $n=2$ we recover our starting ansatz of section 2.

Substituting this ansatz into the reflection equation, we realize that the
simplest possible solution is the symmetric one, namely
\begin{equation}
Y_1(\lambda)=Y_2(\lambda)= \cdots = Y_{n-1}(\lambda)~~~{\rm and}~~~
Y_{n+2}(\lambda)=Y_{n+1}(\lambda)= \cdots = Y_{2n}(\lambda) 
\end{equation}

It turns out that the remaining functional equations for the functions $Y_1(\lambda)$, $Y_n(\lambda)$,
$Y_{n+1}(\lambda)$, $Y_{2n}(\lambda)$, $Y_{2n+1}(\lambda)$ and $Y_{2n+2}(\lambda)$ are very similar
to those presented in the appendix A. Therefore, they can be solved by the same procedure described
in section 2 and in what follows we only quote our final results. As before we find three general families
of $K$-matrices, and the diagonal one is given by
\begin{equation}
Y_1^{(1)}(\lambda ) = 1,~~
Y_n^{(1)}(\lambda ) = \frac{e^{\lambda }-I{q}^{-(n-1)/2}}{e^{-\lambda }-I {q}^{-(n-1)/2}}
\end{equation}
\begin{equation}
Y_{n+1}^{(1)}(\lambda ) = \frac{e^{\lambda }+I{q}^{-(n-1)/2}}{e^{-\lambda }+I {q}^{-(n-1)/2}},~~
Y_{2n}^{(1)}(\lambda ) = \frac{e^{\lambda }+I{q}^{-(n-1)/2}}{e^{-\lambda }+I {q}^{-(n-1)/2}}
\frac{e^{\lambda }-I{q}^{(n-1)/2}}{e^{-\lambda }-I {q}^{(n-1)/2}}
\end{equation}

The one-parameter families of non-diagonal $K$-matrices are given by
\begin{equation}
Y_1^{(2)}(\lambda,\xi_-)  =  (e^{2\lambda }+q^{n-1})(\xi _-^2q^{n-1}e^{2\lambda }-1 )e^{-\lambda},~~
Y_{2n}^{(2)}(\lambda,\xi_-)  =  (e^{2\lambda }+q^{n-1})(\xi _-^2q^{n-1}-e^{2\lambda })e^{\lambda }
\end{equation}
\begin{equation}
Y_{n}^{(2)}(\lambda,\xi_-)  =  \frac{(1+e^{2\lambda })}{2}\left[2(e^{2\lambda }-1)\xi _-q^{n-1}-e^{\lambda }(1+q^{n-1})(1-\xi _-^2q^{n-1})\right]
\end{equation}
\begin{equation}
Y_{n+1}^{(2)}(\lambda,\xi_-)  =  \frac{(1+e^{2\lambda })}{2}\left[-2(e^{2\lambda }-1)\xi _-q^{n-1}-e^{\lambda }(1+q^{n-1})(1-\xi _-^2q^{n-1})\right]
\end{equation}
\begin{equation}
Y_{2n+1}^{(2)}(\lambda,\xi_-)=Y_{2n+2}^{(2)}(\lambda,\xi_{-})  =  \frac{(e^{2\lambda }-1)}{2}(1-q^{n-1})(\xi _-^2q^{n-1}+1)e^{\lambda }
\end{equation}
and
\begin{equation}
Y_1^{(3)}(\lambda,\xi_-)  =  (e^{2\lambda }-q^{n-1})(\xi_{-}e^{2\lambda }-1 )e^{-\lambda},~~
Y_{2n}^{(3)}(\lambda,\xi_-)  =  (e^{2\lambda }-q^{n-1})(\xi_{-}-e^{2\lambda })e^{\lambda }
\end{equation}
\begin{equation}
Y_{n}^{(3)}(\lambda,\xi_-)=Y_{n+1}^{(3)}(\lambda,\xi_{-})  =  \frac{(1+e^{2\lambda })}{2}(1-q^{n-1})(\xi_{-}-1)e^{\lambda }
\end{equation}
\begin{equation}
Y_{2n+1}^{(3)}(\lambda,\xi_-)  =   \frac{(e^{2\lambda }-1)}{2}\left[2(e^{2\lambda }+1)\sqrt{\xi_{-}}q^{(n-1)/2}+(1+q^{n-1})(1+\xi _-)e^{\lambda }\right]
\end{equation}
\begin{equation}
Y_{2n+2}^{(3)}(\lambda,\xi_-)  =   \frac{(e^{2\lambda }-1)}{2}\left[-2(e^{2\lambda }+1)\sqrt{\xi_-}q^{(n-1)/2}+(1+q^{n-1})(1+\xi _-)e^{\lambda }\right]
\end{equation}

The next natural step is to search for asymmetric $K$-matrices for $n \geq 3$, i.e. those having
$Y_1(\lambda) \neq Y_2(\lambda) \neq \cdots Y_{n-1}(\lambda)$ and 
$Y_{n+2}(\lambda) \neq Y_{n+3}(\lambda) \neq \cdots Y_{2n}(\lambda)$. In this case the number
of free parameters grows rapidly with $n$ and the solution of the reflection equation becomes more
involving. To illustrate that, we consider the $D_3^2$ model and for sake
of simplicity we look first for diagonal solutions. There are six functions $Y_j(\lambda)$ to be 
determined and their ratios are fixed by choosing some easy looking relations coming from
the reflection relation. More precisely, we have found the following equations
\begin{equation}
\frac{Y_2(\lambda )}{Y_1(\lambda)} =\frac{e^{2\lambda}-c_1}{e^{-2\lambda}-c_1},~~
\frac{Y_3(\lambda )}{Y_1(\lambda)} =\frac{e^{\lambda}-c_2}{e^{-\lambda}-c_2},~~
\frac{Y_4(\lambda )}{Y_1(\lambda)} =\frac{e^{\lambda}-c_3}{e^{-\lambda}-c_3}
\end{equation}
\begin{equation}
\frac{Y_6(\lambda )}{Y_4(\lambda)} =\frac{e^{\lambda}-c_4}{e^{-\lambda}-c_4},~~
\frac{Y_5(\lambda )}{Y_6(\lambda)} =\frac{e^{-2\lambda}-c_5}{e^{2\lambda}-c_5}
\end{equation}
where $c_j$ are once again constants yet to be determined. Substituting 
these relations back to the reflection equation
we find only one possible manifold for the parameters $c_j$, which reads 
\begin{equation}
c_1=c_5=-1~~{\rm and}~~c_2=c_3=c_4=0
\end{equation}

After an appropriate normalization, this solution leads us to a new diagonal $K$-matrix for the
$D_3^2$ model
\begin{equation}
K_{-}^{D_3^2}(\lambda)={\rm diag}(e^{-2\lambda},1,1,1,1,e^{2\lambda})
\end{equation}

It  is plausible that this ``almost unity'' 
solution and its extensions generalizes for arbitrary values of $n \geq 4$.
Next we have looked at the possibility of asymmetric non-diagonal solutions for the $D_3^2$ model. It turns out that,
within our algebraic approach, we did not found any of such solutions. However, this possibility should not be
completely rule out, at least for general $n$, since we have so many free parameters that the chance to miss
a particular integrable manifold is high. In general,
classification of the solutions of the reflection equation seems to be an intricated problem even
for simpler models \cite{CLA}. We hope, however, that our $K$-matrices results prompt 
further investigation concerning
this problem for the
$D_n^2$ vertex models. 

We would like to conclude this section with the following remarks. The
$K_{+}(\lambda)$ matrices can be obtained from $K_{-}(\lambda)$ by the
isomorphism
\begin{equation}
K_{+}(\lambda)=K_{-}^{t}[(n-1)\ln[q]-\lambda]M
\end{equation}
where $M$ is a diagonal matrix given by
\begin{equation}
M= {\rm diag}(q^{(2n-3)},q^{2n-5}, \cdots, 1,1, \cdots, q^{-(2n-5)}, q^{-(2n-3)})
\end{equation}

Once we are equipped with $K_{\pm}(\lambda)$ matrices, the construction of
the corresponding open spin chains is possible along the lines of section 2. Similarly,
at least for the diagonal
solution, one can also repeat our
Bethe ansatz construction without further 
technical difficulties. 
In particular, we conjecture
that the open spin chain associated to the first non-diagonal solution at 
$\xi_{\pm}=0$ is the one having the 
underlying quantum group
symmetry.

\section{Concluding Remarks}
In this paper we have made a great deal of progress towards the understanding of the integrability
of the $D_n^2$ vertex model  with open boundaries.
We have investigated the solutions of the associated reflection equation 
and  found three general families of $K$-matrices which respect the minimal
$U(1)$ symmetries of this system. We have carried out a Bethe ansatz analysis for the simplest case,
$D_2^2$ model, revealing to us that the first non-diagonal solution at $\xi_{\pm}=0$ possesses 
the special quantum algebra symmetry. In fact, the structure of the $K$-matrices at this 
particular point leads us to
conjecture that this will be the case for arbitrary values of $n$.

We believe that our results open an enormous avenue for further investigations. One clear
possibility is to use the Bethe ansatz results
of section 3 to compute the thermodynamic behaviour, the bulk and the surface critical
exponents.  
It would be also interesting to generalize our results of section 3 for all sort of open
boundary conditions and for arbitrary values of $n$. 
In this case, probably the most suitable tool would be instead
the algebraic Bethe ansatz approach. This method would allow us to show that indeed the
Bethe ansatz states are highest weight states of the underlying quantum algebra in the
case of the first non-diagonal family at $\xi_{\pm}=0$.

Other interesting issue is to apply the notion of 
the quantum group twisting \cite{RE1} to find out slightly different
$D_n^2$ $R$-matrices. As a result, this might lead us to 
integrable models with very different behaviour, for an example see ref.\cite{AN}. The practical
implementation of twisting, however, seems to be quite involving 
specially for an algebra such $D_n^2$. To shed
some light to this problem we proceed in a much more phenomenological way. Motived by the structure of
the non-diagonal solutions,
we add extra Boltzmann weights to the Jimbo's $R$-matrix to account for such boundary terms at the level of the
associated bulk Hamiltonian. Next step is to try to solve the Yang-Baxter equation for this novel $R$-matrix structure.
It turns out that we succeed to find a new $R$-matrix solution for the $D_2^2$ model. Since this involves many
technicalities, we have summarized it in appendix B together with the study of the corresponding solutions
of the reflection equation. We hope that these results will be useful to motivate
further progress in this
problem.

\section*{Acknowledgements} The work of M.J. Martins has been partially supported by the Lampadia Foundation
and by the Brazilian research Agencies CNPq and Fapesp. X-W. Guan thanks Fapesp  
and DFG-SFB393  
for financial support and the hospitality  at the
the Institut f\"{u}r Physik, Technische Universit\"{a}t, Chemnitz .

\centerline{\bf Appendix A : $R$ matrix properties and reflection equations}
\setcounter{equation}{0}
\renewcommand{\theequation}{A.\arabic{equation}}

In this appendix we briefly discuss some useful properties of the $D^2_n $ $R$-matrix
\cite{JI}. We also present for $n=2$ some relevant relations derived from
the reflection equation.
The $D^2_n $ $R$-matrix satisfies, besides the unitarity and regularity, extra
relations denominated PT invariance and crossing symmetry, 
PT-Symmetry:
\begin{equation}
P_{12}R_{12}(\lambda )P_{12}=R^{t_1t_2}_{12}(\lambda )
\end{equation}
Crossing-symmetry:
\begin{equation}
R_{12}(\lambda )=\frac{\zeta [\lambda ]}{\zeta [(n-1)\ln [q]-\lambda ]}\stackrel{1}{V}R_{12}^{t_2}[(n-1)\ln [q]-\lambda )\stackrel{1}{V}^{-1}
\end{equation}
where $\zeta (\lambda )$ is a normalization function and $V$ is the following crossing matrix
\begin{equation}
V={\rm antidiag}\left\{q^{-(2n-3)/2},q^{-(2n-5)/2},\cdots,  \frac{1}{\sqrt{q}},1,1,
\sqrt{q},\cdots, q^{(2n-5)/2},q^{(2n-3)/2}\right\}.
\end{equation}

Here we find convenient to normalize the original Jimbo's $R$-matrix by an overall
factor $e^{2\lambda }q^n$ and the function $\zeta (\lambda )$ is given by
\begin{equation}
\zeta (\lambda )=(e^{\lambda }-e^{-\lambda  })(\frac{e^{\lambda }}{q^{(n-1)}}-
\frac{q^{(n-1)}}{e^{\lambda }})
\end{equation}

Next we present the simplest relations derived from the reflection equation we used
in section 2. For sake of simplicity we shall use the following notation
$Y_i(x)\equiv Y_i,Y_i(y)\equiv Y_i^{'}, w_j(x-y)\equiv w_j, w_j(x+y)=w_j^{'}$.
Considering this notation, the relations we have selected from the reflection
equation are given by
\begin{eqnarray}
-w^{'}_2\left[-w_3Y_1^{'}Y_2+w_4Y_1Y_2^{'}-w_5Y_1^{'}Y_5+w_6Y_1Y_6^{'}\right]  =\nonumber\\ 
w^{}_2\left[-w_4^{'}Y_1^{'}Y_1+w_5^{'}(Y_2^{'}Y_5^{}+Y_2^{}Y_6{'})+w_3^{'}(Y_2^{'}Y_2^{}+Y_5^{}Y_6{'})\right] \label{A.1}\\
-w^{'}_2\left[-w_3Y_1^{'}Y_3+w_4Y_1Y_3^{'}+w_6Y_1^{}Y_5^{'}-w_5Y_1^{'}Y_6^{}\right] = \nonumber\\ 
w^{}_2\left[-w_4^{'}Y_1^{'}Y_1+w_5^{'}(Y_3^{}Y_5^{'}+Y_3^{'}Y_6{})+w_3^{'}(Y_3^{'}Y_3^{}+Y_5^{'}Y_6{})\right] \label{A.2}\\
-w^{'}_2\left[w_3Y_3^{'}Y_4-w_4Y_3Y_4^{'}+w_5Y_4^{}Y_5^{'}-w_6Y_4^{'}Y_6^{}\right]  = \nonumber\\ 
w^{}_2\left[-w_3^{'}Y_4^{'}Y_4+w_6^{'}(Y_3^{}Y_5^{'}+Y_3^{'}Y_6{})+w_4^{'}(Y_3^{'}Y_3^{}+Y_5^{'}Y_6{})\right] \label{A.3}\\
-w^{'}_5\left\{w_5(Y_6^{'}Y_6-Y_5Y_5^{'})+w_3\left[Y_2^{'}(Y_5^{}-Y_6^{})+Y_2(Y_6^{'}-Y_5^{'})\right]\right\}  = \nonumber\\
w^{'}_3\left\{w_3(Y_6^{'}Y_5-Y_6Y_5^{'})+w_5\left[Y_2^{'}(Y_6^{}-Y_5^{})+Y_3(Y_6^{'}-Y_5^{'})\right]\right\} \label{A.4}\\
-w^{'}_3\left\{w_5(Y_3^{'}Y_3-Y_2Y_2^{'})+w_3\left[Y_5^{}(Y_3^{'}-Y_2^{'})+Y_5^{'}(Y_2^{}-Y_3^{})\right]\right\}  = \nonumber\\
w^{'}_5\left\{w_3(Y_2^{}Y_3^{'}-Y_3Y_2^{'})+w_5\left[Y_5^{'}(Y_3^{}-Y_2^{})+Y_6(Y_3^{'}-Y_2^{'})\right]\right\} \label{A.5}\\
-w^{'}_6\left\{w_6(Y_6^{'}Y_6-Y_5Y_5^{'})+w_4\left[Y_2^{'}(Y_5^{}-Y_6^{})+Y_2(Y_6^{'}-Y_5^{'})\right]\right\}  = \nonumber\\
w^{'}_4\left\{w_4(Y_6^{'}Y_5-Y_6Y_5^{'})+w_6\left[Y_2^{'}(Y_6^{}-Y_5^{})+Y_3(Y_6^{'}-Y_5^{'})\right]\right\} \label{A.6}\\
-w^{'}_4\left\{w_6(Y_3^{'}Y_3-Y_2Y_2^{'})+w_4\left[Y_5^{}(Y_3^{'}-Y_2^{'})+Y_5^{'}(Y_2^{}-Y_3^{})\right]\right\}  = \nonumber\\
w^{'}_6\left\{w_4(Y_3^{'}Y_2-Y_3Y_2^{'})+w_6\left[Y_5^{'}(Y_3^{}-Y_2^{})+Y_6(Y_3^{'}-Y_2^{'})\right]\right\} \label{A.7}
\end{eqnarray}

The functions $w_j(\lambda )$ are some of the Boltzmann weights of $D^2_2$ model and are given by
\cite{JI}
\begin{equation}
w_2(\lambda )= 
(e^{\lambda }-e^{-\lambda })(\frac{e^{\lambda }}{q}-\frac{q}{e^{\lambda }}),~~
w_3(\lambda )  = 
-\frac{1}{2}(q-\frac{1}{q})(\frac{e^{\lambda }}{q}-\frac{q}{e^{\lambda }})(e^{-\lambda }+1)
\end{equation}
\begin{equation}
w_4(\lambda )  = 
-\frac{1}{2}(q-\frac{1}{q})(\frac{e^{\lambda }}{q}-\frac{q}{e^{\lambda }})(e^{\lambda }+1),~~
w_5(\lambda )  = 
\frac{1}{2}(q-\frac{1}{q})(\frac{e^{\lambda }}{q}-\frac{q}{e^{\lambda }})(-e^{\lambda }+1)
\end{equation}
\begin{equation}
w_6(\lambda )  = 
-\frac{1}{2}(q-\frac{1}{q})(\frac{e^{\lambda }}{q}-\frac{q}{e^{\lambda }})(e^{\lambda }-1)
\end{equation}

\centerline{\bf Appendix B : A new $D_2^2$ $R$-matrix }
\setcounter{equation}{0}
\renewcommand{\theequation}{B.\arabic{equation}}

We begin by presenting the new $D_2^2$ $R$-matrix
\begin{eqnarray}
R(\lambda ) & = & (e^{2\lambda}-q^2)^2 \sum_{\alpha \neq 2,3}
E_{\alpha \alpha }
\otimes E_{\alpha \alpha }
+q(e^{2\lambda}-1)(e^{2\lambda}-q^2)
\sum _{\mbox{\scriptsize $\begin{array}{c}\alpha \neq \beta,\beta^{'}\\
\alpha {\rm or} \beta \neq 2,3 \end{array}$}}
E_{\alpha \alpha }\otimes E_{\beta \beta }  \nonumber \\
&& 
-\frac{(q^2-1)(e^{2\lambda}-q^2)}{2} [
(e^{\lambda}+1)
\sum _{\mbox{\scriptsize $\begin{array}{c}\alpha<2\\
\beta=2,3 \end{array}$}} + e^{\lambda}(e^{\lambda}+1)
\sum _{\mbox{\scriptsize $\begin{array}{c}\alpha>3 \\  \beta=2,3
\end{array}$}} ] (E_{\alpha\beta} \otimes E_{\beta\alpha}+
E_{\beta^{'}\alpha^{'}} \otimes E_{\alpha^{'}\beta^{'}}) \nonumber \\
&& 
-\frac{(q^2-1)(e^{2\lambda}-q^2)}{2}  [
(1-e^{\lambda})
\sum _{\mbox{\scriptsize $\begin{array}{c}\alpha<2\\
\beta=2,3 \end{array}$}} + e^{\lambda}(e^{\lambda}-1)
\sum _{\mbox{\scriptsize $\begin{array}{c}\alpha>3\\ \beta=2,3 
\end{array}$}} ] (E_{\alpha\beta} \otimes E_{\beta^{'}\alpha}+
E_{\beta^{'}\alpha^{'}} \otimes E_{\alpha^{'}\beta}) \nonumber \\
& &
+\sum _{\alpha,\beta \neq 2,3 }a_{\alpha \beta }(\lambda)
E_{\alpha \beta }\otimes E_{\alpha ^{'}\beta ^{'}} \nonumber \\
&& 
+\sum _{\mbox{\scriptsize $\begin{array}{c}\alpha \neq 2,3\\
\beta=2,3 \end{array}$}} 
b_{\alpha}^{+}(\lambda)
E_{\alpha\beta} \otimes E_{\alpha{'}\beta^{'}}+
{\tilde{b}}_{\alpha}^{+}(\lambda)
E_{\beta^{'}\alpha^{'}} \otimes E_{\beta\alpha}+
b_{\alpha}^{-}(\lambda)
E_{\alpha\beta} \otimes E_{\alpha{'}\beta}+
{\tilde{b}}_{\alpha}^{-}(\lambda)
E_{\beta\alpha^{'}} \otimes E_{\beta\alpha}  \nonumber \\
&&
+\sum _{\alpha= 2,3 }c^{+}(\lambda)
E_{\alpha\alpha} \otimes E_{\alpha^{'}\alpha^{'}}  
+c^{-}(\lambda)
E_{\alpha\alpha} \otimes E_{\alpha\alpha}  
+d^{+}(\lambda)
E_{\alpha\alpha^{'}} \otimes E_{\alpha^{'}\alpha}  
+d^{-}(\lambda)
E_{\alpha\alpha^{'}} \otimes E_{\alpha\alpha^{'}}   \nonumber \\
&&
+\sum _{\alpha= 2,3 }f(\lambda) \left [
E_{\alpha\alpha^{'}} \otimes E_{\alpha\alpha}   
+E_{\alpha\alpha} \otimes E_{\alpha\alpha^{'}}   
-E_{\alpha\alpha} \otimes E_{\alpha^{'}\alpha}   
-E_{\alpha^{'}\alpha} \otimes E_{\alpha\alpha}    \right ] \nonumber \\
\end{eqnarray}
where $E_{\alpha \beta}$
are the elementary $4\times 4$ matrices and we set
$\alpha ^{'}=5-\alpha $. The  Boltzmann weights  are given  by
\begin{equation}
a_{11}(\lambda)=a_{44}(\lambda)=q^2(e^{2\lambda}-1)^2,~~a_{14}(\lambda)=a_{41}(\lambda)e^{-2\lambda}=
(q-1)(q^2-1)(e^{2\lambda}+q)
\end{equation}
\begin{equation}
b_1^{\pm } = \pm \frac{q^{3/2}}{2}(q^2-1)(e^{2\lambda }-1)(e^{\lambda} \pm1 ),~~
{\tilde{b}}_1^{\pm } = \pm \frac{q^{-1/2}}{2}(q^2-1)(e^{2\lambda }-1)(e^{\lambda } \pm q^2)
\end{equation}
\begin{equation}
b_4^{\pm } =  \frac{q^{1/2}}{2}e^{\lambda}(q^2-1)(e^{2\lambda}-1)(e^{\lambda } \pm q^2),~~
{\tilde{b}}_4^{\pm } = \frac{q^{1/2}}{2}e^{\lambda}(q^2-1)(e^{2\lambda}-1)(e^{\lambda } \pm 1)
\end{equation}
\begin{equation}
d^{\pm} = \pm\frac{e^{\lambda}}{4}(q^2-1)^2(e^{\lambda }\pm 1)^2,~~
f(\lambda)=\frac{e^{\lambda}}{4} (e^{2\lambda}-1)(q^2-1)^2
\end{equation}
\begin{equation}
c^{\pm} =  \pm\frac{e^{\lambda}}{4}(q^2-1)(e^{\lambda}\mp 1)[e^{\lambda}(3+q^2) \pm(1+3q^2)] 
+q(e^{2\lambda }-1)(e^{2\lambda}-q^2)
\end{equation}

This $R$-matrix has additional Boltzmann weights, the last term in equation (B.1), 
as compare to the
standard $D_2^2$ $R$-matrix \cite{JI}. In addition, several other
weights have also a different functional dependence on the spectral paramater $\lambda$.
For periodic boundary conditions,
such differences are not important since we verified, by using the algebraic Bethe ansatz 
approach \cite{MAR}, that the corresponding Bethe ansatz equations and eigenvalues are the same
as those found for the standard $D_2^2$ model \cite{RE,MAR}. This result is a strong indication
that indeed the  $R$-matrix (B.1) can be obtained by twisting the usual $D_2^2$ $R$-matrix.
However, the situation for open boundary conditions turns out to be a bit different.  In fact,
we did not find any diagonal solution of the corresponding reflection equation. The
basic $K$-matrices are non-diagonal and we managed to find two classes of such solutions.
The first family depends only on a discrete parameter $\varepsilon=\pm$ and is given by
\begin{eqnarray}
Y_1^{(1,\varepsilon)}(\lambda,\xi_-) & = & Y_4^{(1,\varepsilon)}(\lambda,\xi_-) =  (e^{2\lambda }+\varepsilon \,q)\\
Y_2^{(1,\varepsilon)}(\lambda,\xi_-) & = & Y_3^{(1,\varepsilon)}(\lambda,\xi_-)  = \frac{1}{2}(1+\varepsilon \,q)(1+e^{2\lambda })\\
Y_5^{(1,\varepsilon)}(\lambda,\xi_-) & = &  Y_6^{(1,\varepsilon)}(\lambda,\xi_-)  = \frac{1}{2}(1-\varepsilon \,q)(-1+e^{2\lambda })
\end{eqnarray} 
while the second family has an extra continuous parameter $\xi_{-}$
\begin{eqnarray}
Y_1^{(2,\varepsilon)}(\lambda,\xi_-) & = & (e^{2\lambda }-\xi _-^2)(q+\varepsilon \,e^{2\lambda })e^{-\lambda },\\
Y_4^{(2,\varepsilon)}(\lambda,\xi_-) & = & (1-\xi _-^2e^{2\lambda })(q+\varepsilon \,e^{2\lambda })e^{\lambda },\\
Y_2^{(2,\varepsilon)}(\lambda,\xi_-) & = & \frac{1}{2}(1-\xi _-e^{\lambda })(\xi _-+e^{\lambda })(1+e^{2\lambda })(\varepsilon +q),\\
Y_3^{(2,\varepsilon)}(\lambda,\xi_-) & = & \frac{1}{2}(1+\xi _-e^{\lambda })(-\xi _-+e^{\lambda })(1+e^{2\lambda })(\varepsilon +q),\\
Y_5^{(2,\varepsilon)}(\lambda,\xi_-) & = &  \frac{1}{2}(e^{2\lambda }-1)(\varepsilon -q)(\xi _--e^{\lambda })(\xi _-e^{\lambda }-1),\\
Y_6^{(2,\varepsilon)}(\lambda,\xi_-) & = &  \frac{1}{2}(e^{2\lambda }-1)(\varepsilon -q)(\xi _-+e^{\lambda })(\xi _-e^{\lambda }+1),
\end{eqnarray} 

Finally, we remark that since this new $R$-matrix is only unitary, the associated $K_{+}(\lambda)$
matrices can not be directly obtained by an isomorphism of the type described in (20).
However, as shown in ref. \cite{CHI}, 
unitarity is a sufficient condition to allow one to construct commutative transfer matrices leading
to open spin chains. In this case one has to solve an extra reflection equation to
obtain the $K_{+}(\lambda)$ matrix \cite{CHI}.

\end{document}